\documentstyle[aps,prb,preprint,eqsecnum,epsf]{revtex}
\begin{document}
\bibliographystyle{unsrt}
\title{A molecular-dynamics study of ductile and brittle fracture in model noncrystalline solids}
\renewcommand{\thefootnote}{\fnsymbol{footnote}}
\author{M.L. Falk\footnote{Present address: Division of Engineering and Applied Sciences, Harvard University, Cambridge, MA 02138; e-mail: {\tt falk@esag.harvard.edu}}}
\renewcommand{\thefootnote}{\arabic{footnote}}
\address{Department of Physics, University of California,
        Santa Barbara, CA 93106}
\date{\today}
\maketitle
\begin{abstract}
Molecular-dynamics simulations of fracture in metallic glass-like
systems are observed to undergo embrittlement due to a small change in
interatomic potential.  This change in fracture toughness, however, is
not accompanied by a corresponding change in flow stress.  Theories of
brittle fracture proposed by Freund and Hutchinson indicate that
strain rate sensitivity is the controling physical parameter in these
cases.  A recent theory of viscoplasticity in this class of solids by
Falk and Langer further suggests that the change in strain rate
sensitivity corresponds to a change in the susceptibility of local
shear transformation zones to applied shear stresses.  A simple model
of these zones is develped in order to quantify the dependence of this
sensitivity on the interparticle potential.
\end{abstract}


\section{Introduction}
This paper presents simulations in which a small change in
interparticle potential leads to a qualitative change in ductility.
Section~\ref{sec:sims} describes the simulation technique and
observations.  Section~\ref{sec:deform} details a technique for
calculating a quantitative measure of local non-affine deformation
which is applicable to materials with no crystalline order, and using
this technique pinpoints those areas of the material which are
undergoing some molecular level rearrangement akin to a dislocation.
Section~\ref{sec:analysis} relates this observed change in ductility
to the particular change in interparticle potential.  This final
section discusses the simulations in terms of current phenomenological
theories of brittle and ductile behavior due to Freund and
Hutchinson\cite{Freund85} and a theory of this mechanism of molecular
level rearrangement in non-crystalline materials developed by Falk and
Langer which we shall refer to as FL.\cite{Falk98} Finally, a
simplified microscopic model is analyzed.  This model directly relates
the interparticle potential to the parameter which controls
deformation in FL and the change of fracture toughness observed in the
simulations.  The model also explains the observations of Srolovitz,
{\it et. al.} relating these regions to ``$\tau$-defects'' in previous
metallic glass simulations.\cite{Srolovitz81,Srolovitz83}

The concepts of brittleness and ductility are central to any
understanding of failure in solids.  The most developed
first-principles theories of ductility are rooted in the dynamics of
dislocations in crystalline solids.\cite{Rice74,Rice92,Zhou94}
Although it has been conjectured that an analog to a crystal
dislocation exists in noncrystalline solids,\cite{Gilman73} it remains
unclear how to make the direct connection to molecular level phenomena
necessary for these theories to be useful in quantitatively
understanding transitions between brittle and ductile behavior in
disordered materials.  In fact, it is not at all clear that a
dislocation model of this sort is the most appropriate way to
understand materials without regular structure although these
materials are observed to undergo similar brittle-ductile transitions
to their crystalline counterparts.  It is conjectured here that
dislocation concepts are not the most natural way to describe
non-crystalline solids and theories of ``shear transformation zones''
(STZ's) as first considered in the literature on metallic
glasses\cite{Spaepen77,Argon79a,Argon79b,Spaepen81,Argon83,Khonik94}
are developed as a natural way to understand some of the basic physics
of brittle versus ductile fracture.

\section{Simulations of brittle and ductile fracture in a noncrystalline solid}
\label{sec:sims}

This section describes a series of molecular-dynamics(MD) simulations
of fracture in a simple, two-dimensional amorphous solid.  While these
simulations are clear examples of change in ductility induced by a
change in interparticle potential the important point here is not
simply to differentiate between brittle and ductile behavior, but
rather to establish a connection between this particular change in the
potential and a change the observed fracture toughness.

From a practical standpoint these studies of brittle and ductile
behavior are relevant in the context of several different disordered
and amorphous materials.  The simulated system is most similar
to metallic glasses which have been observed to undergo transitions
between ductile and brittle behavior both as a function of temperature
and due to small amounts of dilute crystallization produced during
annealing.\cite{Pampillo75,Wu90,Johnson96,Gilbert97} 
Similar transitions are also
critical to the processing of colloidal ceramic systems.  These
clay-like materials undergo brittle-ductile transitions due to changes
in salt content, i.e. changes in interparticle
interactions.\cite{Franks96} Issues of brittleness and ductility are
also crucial for the production of high-strength polycrystalline
metallic alloys in which such transitions have been studied
experimentally with respect to temperature and loading
rate.\cite{Noebe92}

In modeling interparticle interactions a Lennard-Jones (LJ) potential
is employed.  This is consistent with previous investigations which
have been carried out in the context of metallic
glasses.\cite{Maeda78,Takeuchi87,Deng89,Srolovitz81,Srolovitz83} While
more sophisticated models of interactions within metals exist many,
such as the Johnson potential for iron\cite{Johnson64} or the model
potential for copper and zirconium employted by Deng, Argon and
Yip,\cite{Deng89} have a basic form similar to that of the LJ
potential.  The use of a simplified potential can be justified here
since these simulations seek to elucidate how a well controlled change
in potential affects fracture behavior rather than exploring the
accuracy of a particular potential.

Despite the simplicity of the LJ potential, the similarity of the
results of these simulations to experiments carried out in metallic
glasses is striking.  First, as in the simulations, both brittle and
ductile fracture are observed in metallic glasses at low temperature
depending on composition.  Typically these glasses display a
pseudo-cleavage fracture mode which involves significant flow at the
crack tip evidenced by vein patterns.\cite{Pampillo75} However,
compositional changes can lead to brittle modes of fracture in which
such flow is not in evidence.\cite{Pampillo74} Some Pd-Si glasses have
even been observed to display the ``mirror'', ``mist'', ``hackle''
behvior typical of the extremely brittle oxide
glasses.\cite{Matsumoto71} Though a transitional temperature to this
brittle mode is seen in many metallic glasses, some have been observed
to experience pseudo-cleavage at all observed temperatures even as low
as 76K.\cite{Pampillo74}

Secondly, as in the simulations, metallic glasses at low temperature
are observed to experience changes in fracture toughness independent
of flow stress.  Tests of the dynamic fracture response and the onset
of flow in Pd-Si and Fe-P-C glasses show a distinct crossover from a
thermally activated to an athermal mechanism for fracture and flow at
temperatures of 273K and 473K respectively.\cite{Chen71,Maddin72} In
the high temperature regime flow and fracture are observed to follow
the same trend when strain rate is varied.  But, in the low
temperature regime analogous to these simulations, the flow stress is
observed to be independent of strain rate while the fracture stress is
strain rate dependent.\cite{Pampillo75} In these simulation a change in
potential produces a change in fracture toughness that is not
accompanied by a corresponding change in the flow stress.  This can be
related theoretically to a change in strain rate sensitivity
independent of the flow stress, and will be discussed in detail in
Section~\ref{sec:analysis}.

\subsection{Methodology}
\label{sec:method}

The simulated systems consisted of 90,000 particles in two-dimensions
interacting via a two-body potential.  In order to avoid problems of
local crystallization, a poly-disperse collection of particles was
simulated.  The system was composed of eight different species in
equal proportion with radii $r_1,r_2,\ldots r_8$ such that 
\begin{equation}
r_\alpha = 1.1\:r_{\alpha-1}; \hspace{1in} \sum_{\alpha=1}^8 \pi r_\alpha^2 = 8 \pi (d_0/2)^2 
\end{equation}
Thus, the total volume of the collection was the same as if the
particles were all of radius $d_0/2$, and the system is roughly
comparable to a single component system in which the rest spacing
between two molecules is $d_0$.  All quantities will be given in terms
of dimensions for which $d_0$ is the length scale.  Therefore in these
units $d_0=1$.  The masses of all particles were taken to be $m = 1$
in these units.

The inter-molecular potential was different in the two simulations.
In the simulation which displayed more ductile behavior the potential was a
standard Lennard-Jones 6-12 potential
\begin{equation}
U^{LJ}_{\alpha \beta}(r) = e \left[\left({r_\alpha+r_\beta \over r}
\right)^{12} - 2 \left({r_\alpha+r_\beta \over r}\right)^{6}\right],
\label{LJpot}
\end{equation}
where $r$ is the interparticle distance and $r_\alpha$ and $r_\beta$
are the radii of the two particles.  $e$, the depth of the energetic
minimum of the two particle interaction, is unity in our units.  Note,
this is not the standard form of writing the LJ potential, which is
typically written in terms of the hard-core radius, $a_\alpha=
2^{-1/6} r_\alpha$.  This expression is, however, equivalent and
will facilitate the introduction of the second potential to which it will be
compared.  

\begin{figure}[h]
\epsfxsize=5.0in
\centerline{\epsfbox{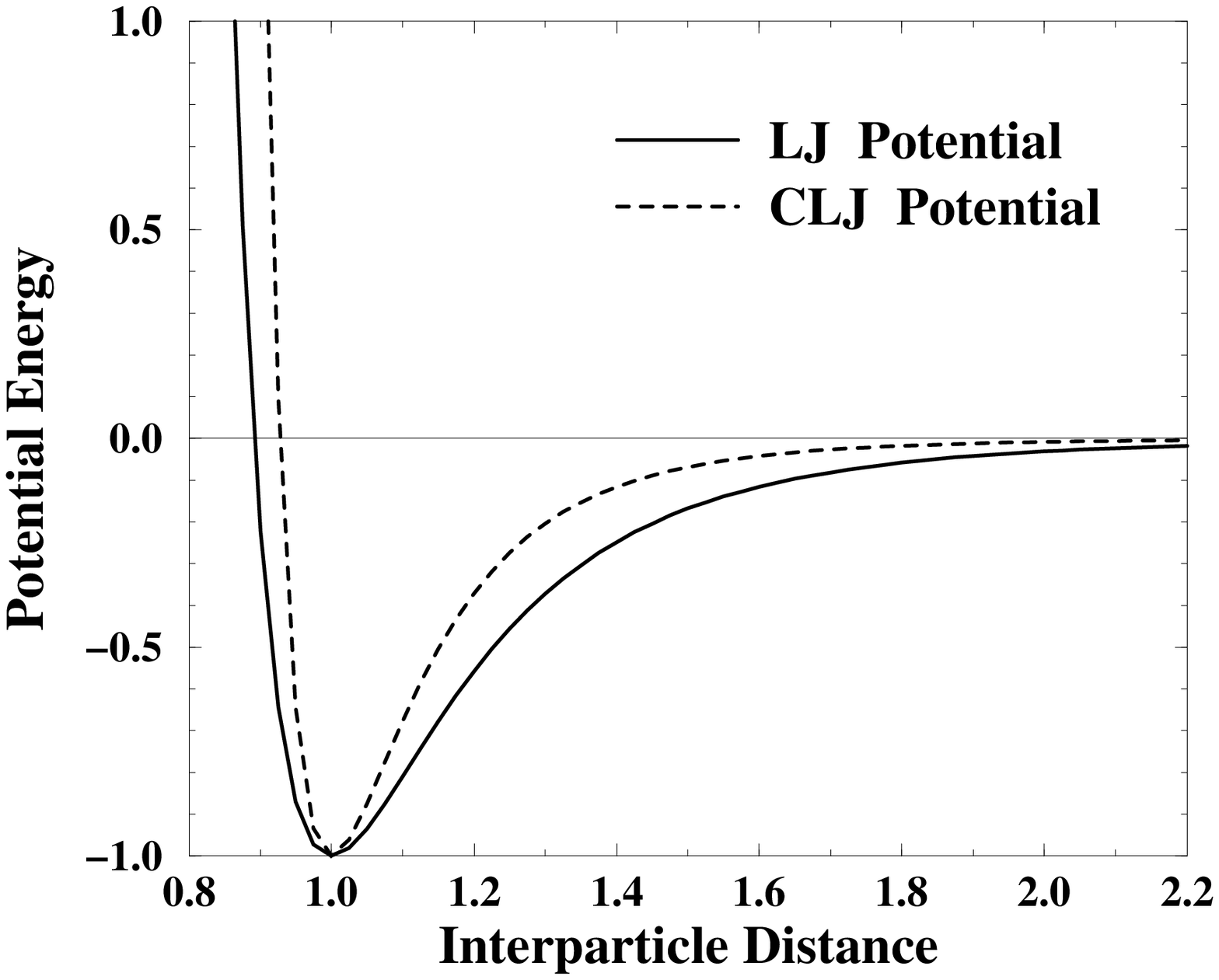}}
\caption{The LJ and CLJ potentials.  Energy is given in units of $e$.
Interparticle distance is given in units of $d_0$.
\label{potentials}}
\end{figure}

The simulation which displayed more brittle behavior employed a
potential which will be refered to as the Compressed
Lennard-Jones(CLJ) potential because it is the Lennard-Jones
potential rescaled around the center of the potential well,
\begin{equation}
U^{CLJ}_{\alpha \beta} (r) = U^{LJ}_{\alpha \beta} (\lambda\:r + (1-\lambda)
(r_\alpha+r_\beta)).
\label{CLJpot}
\end{equation}
Note that the standard LJ potential is recovered when
$\lambda=1$.  Furthermore, $\lambda \rightarrow0$ corresponds to a
mean field limit in which every particle interacts with every other
particle equivalently, and $\lambda \gg 1$ is the limit
of solely nearest neighbor interactions.  For the second simulation
the parameter $\lambda$ was chosen to be 1.5.  This means that width
of the potential well was smaller by 33\%, and, consequently, the
effective range of interaction was also shortened compared to the
standard LJ interaction.  For the sake of comparison
Fig.~\ref{potentials} shows both potentials.  In both cases
interactions were cut off at a range of $d_c \approx 2.2 d_0$.

All times are given in units of $t_0 = d_0 \sqrt{m/e}$.  This
unit of time is approximately equivalent to one molecular vibrational period.

The initial amorphous systems were created by taking 10,000-molecule systems
and equilibrating them using a sequential MD
algorithm with periodic boundary conditions, a Nose-Hoover thermostat
\cite{Nose84,Nose84b,Nose86} and Parrinello-Rahman barostat. 
\cite{Parrinello81,Parrinello82}  The equations of motion and time constants
for the thermostat and barostat were the same as those in
FL.\cite{Falk98} The time step in the simulations was taken to be $0.01
t_0$. The systems were held at low temperature $kT = 0.01 e$ for 5000
timesteps at zero pressure, then the pressure was raised to $10
e/d_0^2$ over the course of 1000 timesteps and lowered again over an
equal period of time.  Subsequently the samples were allowed to
equilibrate at zero pressure for 1000 timesteps.  This procedure
created close-packed samples.  The LJ sample was observed to have
a Young's modulus of 34 and a shear modulus of 10; the CLJ sample
was observed to have a Young's Modulus of 39 and a shear modulus of
12.  These 10,000-molecule samples were then used to create larger
systems by replicating the small system in a $3\times 3$ array.

The larger systems were simulated via a parallel MD algorithm based on
a spatial decomposition method.\cite{Plimpton95} In order to create
the initial conditions for the fracture simulations, the large system
was equilibrated for 100 time steps while held at a very low
temperature, $kT = 0.001 e$.  A crack was then introduced into the
sample.  This was accomplished by imposing displacements as determined
by the analytical solution for a straight crack in an elastic medium
loaded below the ideal critical stress.
The faces of the crack were marked so that the top face would not
interact with the bottom face to prevent the crack from healing.  The
outer boundaries of the system were held fixed while the simulation
was again run for 20,000 timesteps holding the temperature constant to
allow the system to relax.

In the fracture simulations no thermostat or barostat was employed. To
drive the crack, an initial velocity gradient was imposed across the
sample, and the top and bottom surfaces were constrained to move apart
vertically such that the side closer to the crack would separate at a
strain rate of $0.0001 t_0^{-1}$ and the side farthest from the crack
would not move apart at all.  The horizontal motion of these surfaces
was unconstrained.  The left and right surfaces were constrained not
to move in the horizontal direction, though their
vertical motion was unconstrained.

A strain rate of $0.0001 t_0^{-1}$ corresponds to a physical strain
rate on the order of $10^{8} s^{-1}$.  While this may seem high
compared to typical laboratory values, the time for the stresses to
equilibrate is equal to the time for a sound wave to traverse the
sample, $\approx 300 t_0$.  The fact that this time multiplied by the
strainrate is $\approx 0.03 \ll 1$ implies that the system was loaded
nearly quasi-statically.  That is to say that the loading rate was
much slower than the elastic response time, although the loading may
not be slow when compared to the time scale for plastic response.  Of
course, if the crack begins to propagate strain rates near the tip may
be significantly higher.

\subsection{Observations}

\begin{figure}[tb]
\epsfxsize=5.0in
\centerline{\epsfbox{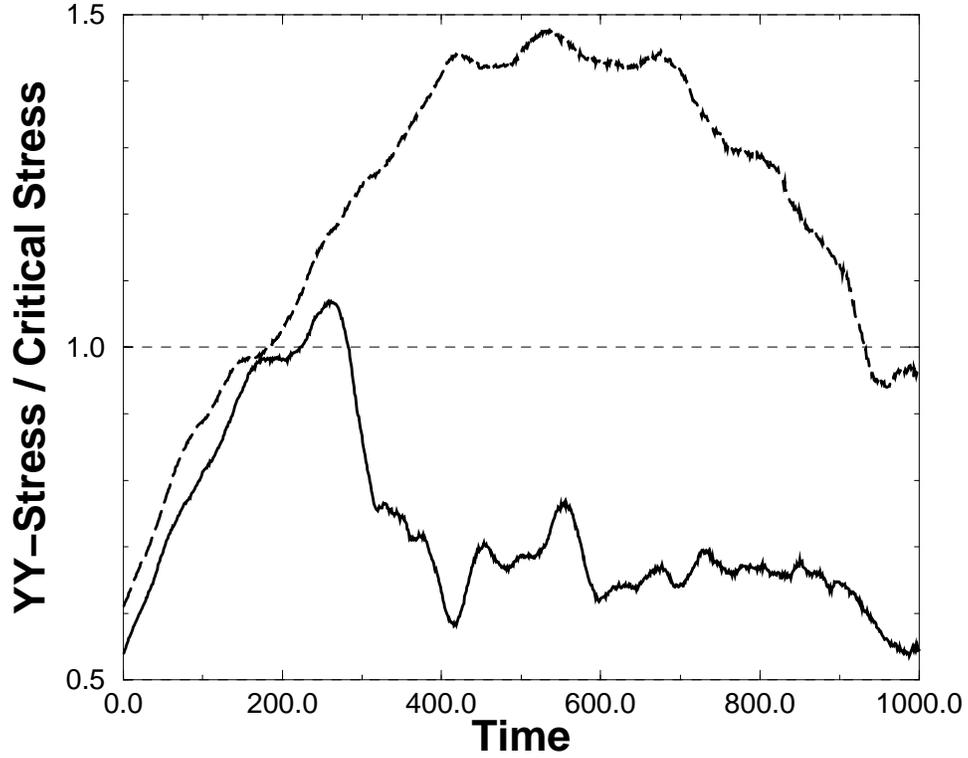}}
\caption{Stress averaged throughout the sample versus time for the CLJ
(solid) and LJ (dashed) simulations.  The higher stress for the onset of fracture in the LJ
case implies increased dissipation.  Stresses are given in units of the
critical stress for failure of an ideally brittle material with the same 
elastic properties.  Time is given in units of $d_0 \sqrt{m / e}$.
\label{graph}}
\end{figure}

Figure \ref{graph} shows the average stress measured during both
simulations.  In order to better compare the two systems, the stresses
are given in units of the critical stress for initial failure of a
perfectly brittle solid with the identical elastic properties,
\begin{equation}
\sigma_c =\sqrt{G E \over \pi a}.
\end{equation}
Here we assume that the system can be treated as a crack in an
infinite medium.  $E$ is Young's Modulus; $a$ is the initial length of
the crack; and $G$ is the energy release rate, which can be expressed
as a surface energy and a dissipation per unit crack extension,
\begin{equation}
G = 2 \gamma + G_{diss}.
\end{equation}
For an ideally brittle solid all the elastic energy released goes into
the creation of new surface, $G_{diss} = 0$.  $\gamma$ was measured by
taking a sample of the material in MD, slicing it along an arbitrary
plane and measuring the change in potential energy.  The value of
$\gamma$ is $1.04 e / d_0$ in the CLJ system and $0.94 e / d_0$ in the
LJ system, thus $\sigma_c^{ideal} = 0.68 e/d_0^2$ in the
ductile system and $0.70 e/d_0^2$ in the brittle system.

Two notable differences are observed between the simulations: (i) In
the CLJ simulation, some modest amount of energy was
dissipated and the crack began to propagate at about 7\% above the
ideal brittle critical stress, but in the LJ simulation fracture
did not proceed until the stress was 48\% above this value.  This
means that for the CLJ case the ratio of energy dissipated to the
energy expended creating surface is 0.14, while for the LJ case
this ratio is 1.19.  (ii) In the CLJ simulation, once the crack
began to propagate, the stress in the system sharply dipped as the
crack moved through the system at speeds reaching 30\% of the shear
wave speed. Throughout this process the crack tip remained atomically
sharp.  The process stopped short of releasing all the stress because
the crack arrested.  In the LJ case, however, the crack tip blunted
significantly.  In this simulation, the stress remained high while
voids nucleated ahead of the tip.  The speed of the ductile crack,
while difficult to measure due to the mechanism of
propagation, stayed well below the speed of the CLJ crack.

\section{Quantifying local deformation}
\label{sec:deform}

The simulation which utilized the CLJ potential resulted in markedly
more brittle behavior than the simulation that utilized the LJ
potential.  In order to address why this particular change in
potential resulted in differing amounts of deformation near the crack
tip the underlying mechanism of deformation must be established.  Work
by Argon and Spaepen suggests that localized deformations, or ``shear
transformation zones,'' are responsible for rearrangements in these
amorphous materials.  This section undertakes an examination of the
microscopic nature of the plastic rearrangement in order to determine
if this is indeed the case for this set of simulations.  These
microscopic observations also serve to differentiate the simulations
performed here from similar investigations of fracture undertaken in
crystals\cite{Zhou96,Abraham94} where plasticity is observed to result
from dislocations emitted from the crack tip or activated in the
vicinity of the crack.

\subsection{Definition of $D^2_{min}$}

In a perfect crystal, dislocations can be readily identified by their
characteristic stress fields or as regions of anomalously high
potential energy.  In glasses, however, such analyses are difficult
due to inhomogeneities frozen into the structure.  Furthermore, it is
not clear that any analog of crystalline dislocations exists in
non-crystalline solids.  For these reasons, a different scheme must be developed to identify regions which deform in a non-affine way and thereby
observe what sort of microscopic structures play the role of
dislocations in these materials.

To identify local rearrangements from a set of molecular positions and
subsequent displacements the closest possible approximation to a local
strain tensor is computed in the neighborhood of any particular
molecule.  The neighborhood is defined in this case by the interaction
range, $d_c$.  The local strain is then determined by minimizing the
mean square difference between the the actual displacements of the
neighboring molecules relative to the central one, and the relative
displacements that they would have if they were in a region of uniform
strain $\varepsilon_{ij}$.  That is, we define
\begin{equation} 
D^2(t, \Delta t) =  \sum_{n}\sum_{i}\Bigl[r^{i}_n(t)-r^{i}_0(t) - 
\sum_{j} (\delta_{ij} + \varepsilon_{ij}) \Bigl(r^{j}_n(t-\Delta t) - 
r^{j}_0(t-\Delta t)\Bigr)
 \Bigr]^2,
\label{Dequation}
\end{equation}
where the indices $i$ and $j$ denote spatial coordinates, and the index 
$n$ runs over the molecules within the interaction range of the 
reference molecule, $n=0$ being the reference molecule. $r^i_n(t)$ is 
the $i$'th component of the position of the $n$'th molecule at time $t$.  
We then find the $\varepsilon_{ij}$ which minimizes $D^2$ by 
calculating: 
\begin{eqnarray} 
X_{ij} & = 
\sum\limits_{n}&(r^{i}_n(t)-r^{i}_0(t)) \times 
(r^{j}_n(t-\Delta t)-r^{j}_0(t-\Delta t)), \\ 
 Y_{ij} & = 
\sum\limits_{n}&(r^{i}_n(t-\Delta t)-r^{i}_0(t-\Delta t)) \times 
(r^{j}_n(t-\Delta t)-r^{j}_0(t-\Delta t)), \\ 
\varepsilon_{ij} & =  \sum\limits_{k}& X_{ik} Y_{jk}^{-1} -
\delta_{ij}. 
\end{eqnarray} 
The minimum value of $D^2(t, \Delta t)$ is then the local deviation
from affine deformation during the time interval $[t-\Delta
t,\,t]$. This quantity shall be refered to as $D_{min}^2$.

\subsection{Molecular Level Observations}
\label{sec:molecobs}

\begin{figure}[p]
\epsfxsize=5.0in
\centerline{\epsfbox{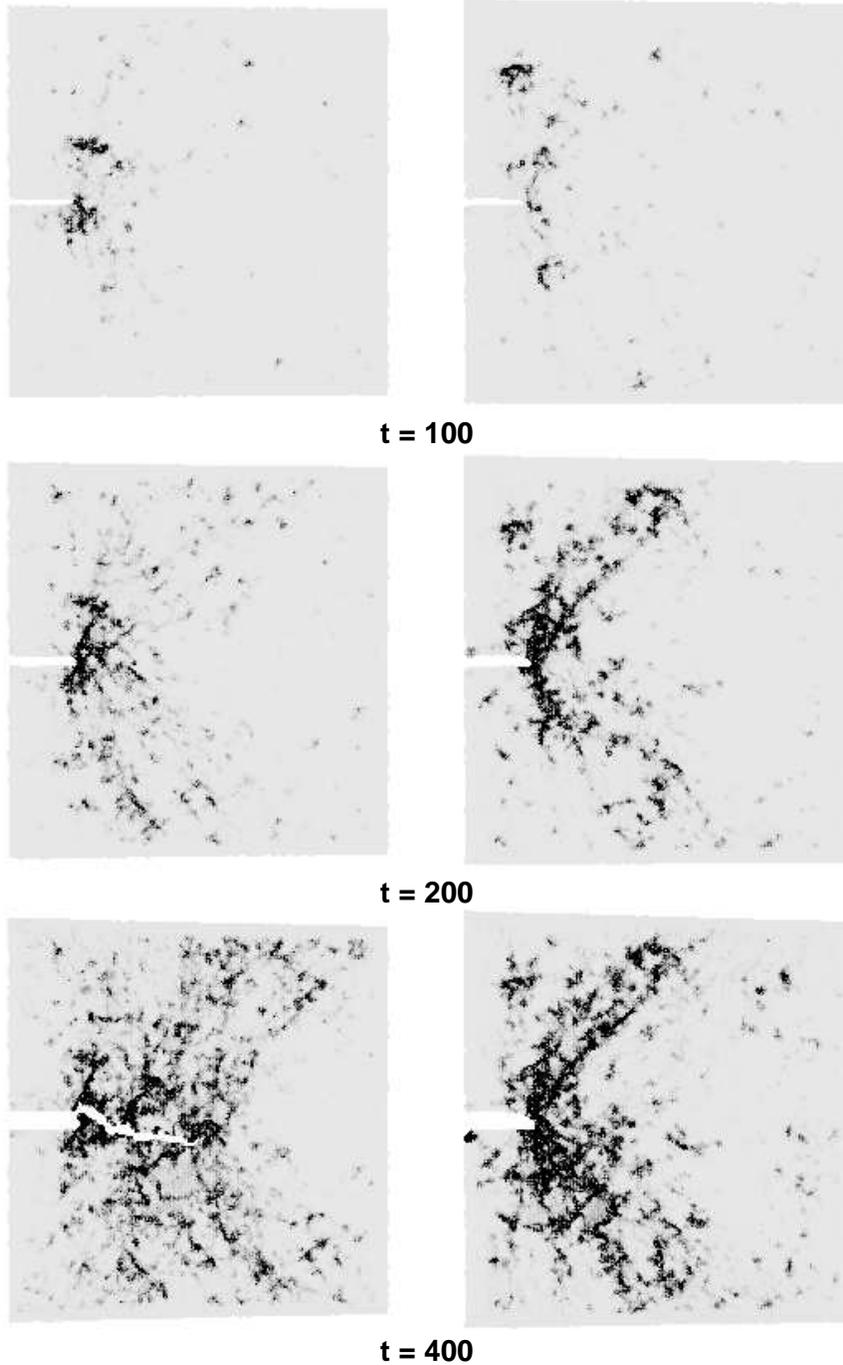}}
\caption{Frames from the CLJ (left) and LJ (right) fracture simulations. In each set the frames are
shaded by the parameter $D^2_{min}$ defined in Eq.~(\ref{Dequation}).  Dark regions
have undergone the highest amount of non-affine rearrangement.  The shading saturates when $D^2_{min}=1$.}
\label{cracksA}
\end{figure}

\begin{figure}[p]
\epsfxsize=5.0in
\centerline{\epsfbox{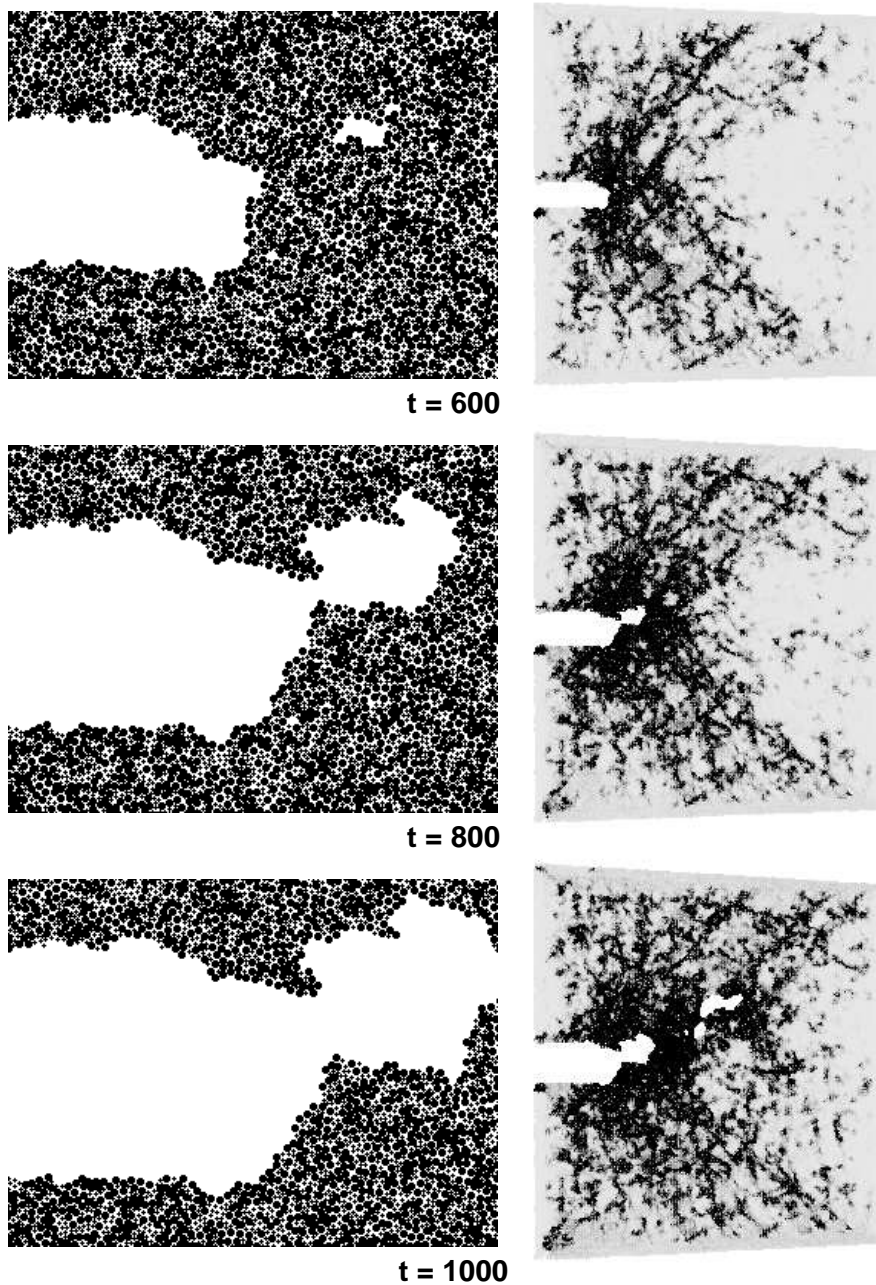}}
\caption{Frames from the LJ simulation showing the nucleation and growth 
of a void in the vicinity of the crack tip.  The frames on the left
are close-ups of the crack tip.  The frames on the right are shaded as
in Figure
\ref{cracksA}.}
\label{cracksB}
\end{figure}

$D_{min}^2$ serves as a diagnostic for
identifying where local rearrangements have taken place.  The
right-hand frames in Figures \ref{cracksA} and \ref{cracksB} are
shaded by the value of $D^2_{min}$ over the interval from $t=0$ to the
current time.  It is immediately apparent that much more non-affine
rearrangement takes place in the LJ simulation than in the
CLJ simulation.  In addition, there seem to develop
preferred directions along which deformation takes place.  These slip
bands which nucleate at the crack tip in the LJ simulation are
clear signs that the dynamics of the plastic response and the
resulting propagating shear modes are crucial aspects of the problem.

\begin{figure}[h]
\epsfxsize=5.0in
\centerline{\epsfbox{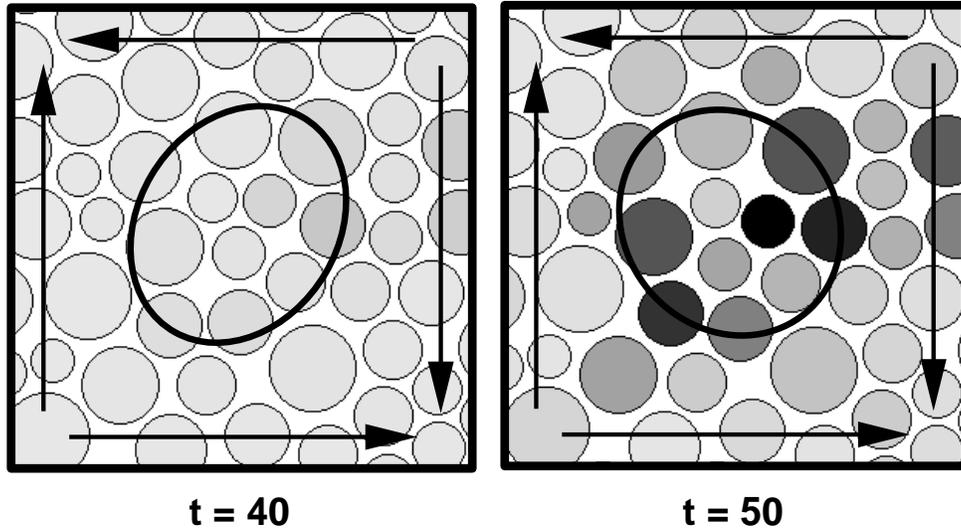}}
\caption{A local region before and after non-affine rearrangement.
The molecules are shaded by $D^2_{min}$, the amount of non-affine
rearrangement.  The arrows denote the approximate direction of the
externally applied shear.  The ovals are included solely as guides for
the eye.}  \label{xform} \end{figure}

Figure \ref{xform} shows one example of a local region before and
after rearrangement.  This rearrangement took place in the early
stages of the ductile simulation prior to significant blunting a small
distance in the y-direction from the tip.  The arrows denote the sense
of the externally applied shear in this region calculated by knowing
the asymptotic stress field near a crack tip.  The figure illustrates
that these regions appear to be of the type discussed by Spaepen as
``flow defects''\cite{Spaepen81} or in other contexts as ``shear
transformation zones''.  That is, the region seems to consist of
roughly 10-20 particles, the rearrangements seem to be local, and the
``defect'' is not mobile in the same sense as a dislocation.
Srolovitz, Maeda, Vitek and Egami established that these rearranging
regions correspond structurally to ``$\tau$-defects,'' regions of
anomalously high local shear stress.\cite{Srolovitz81,Srolovitz83} The
following section will further explore why these regions are
``$\tau$-defects'' and how both the high local stresses and
deformation dynamics arise from the particulars of the intermolecular
potentials.

\section{Analysis}
\label{sec:analysis}

These simulations beg the question as to why this change in
interatomic potential leads to the observed change in ductility.
Unfortunately, the current lack of a detailed understanding of the
microscopics of plasticity in non-crystalline materials will rule out
a detailed first-principles theory of such transitions at this stage.
However, some important connections can be made between these
simulations and current theories of dynamic fracture and the theory of
viscoplasticity in amorphous solids presented in FL.\cite{Falk98} In
addition, toward the end of this section a simplified model of the
molecular rearrangements at the heart of the viscoplasticity theory
will be detailed. This model serves to illustrate how a
first-principles theory may eventually be developed.

\subsection{Macroscopic: brittle-ductile behavior}
We begin by considering the theory of high strain-rate crack growth
proposed by Freund and Hutchinson.\cite{Freund85}  In this theory the
plastic strain rate is considered negligible below some shear stress
$\sigma_{flow}$ and above this stress the strain rate
$\dot{\varepsilon}^{pl}_s$ rises linearly.
\begin{equation} 
\dot{\varepsilon}^{pl}_s = \dot{\varepsilon}_t + \dot{\varepsilon}_0 (\sigma_s - \sigma_{flow})/\mu
\label{plast}
\end{equation}
Here $\mu$ is the shear modulus, $\sigma_s$ is the applied shear
stress, $\dot{\varepsilon}_t$ is the flow rate at yield and $\dot{\varepsilon}_0$ characterizes the strain rate sensitivity.  Furthermore $\dot{\varepsilon}_t \ll
\dot{\varepsilon}_0$ and the effect of $\dot{\varepsilon}_t$ will not
be important for the purpose of this analysis.  Using an assumption of
strain rate dominance, the theory finds that the energy release rate
of the crack is velocity dependent.  Furthermore, the energy release
rate of the brittle crack diverges at both high and low velocity.
Between these two diverging limits there exists a velocity at which
the energy release rate of the crack is a minimum.  According to this
model the crack cannot propagate when driven at less than this minimum
energy release rate.  The value of the minimum energy release rate
depends on the specifics of the plastic response described in
Eq. (\ref{plast}) and, by Freund and Hutchinson's analysis
\begin{equation}
{G_{min} \over G^c_{tip}} \approx  1 + {\cal C} {\dot\varepsilon_0 \over
\sigma_{flow}^2},
\label{Geq}
\end{equation}
where $G^c_{tip}$ is the bare fracture toughness near the tip, and
${\cal C}$ is a proportionality constant which depends on the shear
modulus, density and $G^c_{tip}$.  [NB: We will ignore a second term
proportional to $\dot\varepsilon_t/\dot\varepsilon_0$ for reasons
discussed above.]

In the context of this theory we can ask what would cause one material
to propagate a brittle crack while another admits a more ductile mode of
failure.  Since a given mode of failure can only result if a
propagating solution exists, we can conjecture that the ductile
failure mode results when the propagating brittle solution becomes,
for some reason, inaccessible.  For brittle behavior to have resulted from
the narrowing of the inter-particle potential then, the minimum energy
release rate should have decreased when the potential well width was
narrowed.  This further implies that the narrowing of the potential
either caused a decrease in $\dot{\varepsilon}_0$, the sensitivity of
the strain rate to a change in applied stress, or an increase in
$\sigma_{flow}$, the critical stress for appreciable plastic flow.

The simplest explaination for the increase in brittleness would be
that the narrowing of the potential raised the critical stress for
plastic flow.  This is not the case.  Bulk measurements of
$\sigma_{flow}$ obtained by simulating the two systems in periodic
boundary conditions with zero applied pressure and a constant applied
shear strain rate reveal no significant differences.  $\sigma_{flow}
\approx 0.4 (e / d_0^2)$ for both systems.  This implies that a change
in the flow stress is not the cause of the embrittlement in the
simulations presented here.

Returning for a moment to the Freund and Hutchinson model, we note
that having eliminated $\sigma_{flow}$ as the responsible parameter
for the change in the mode of failure, we must consider the parameter
$\dot{\varepsilon}_0$.  This parameter corresponds to the sensitivity
of the strain rate to an applied stress above the flow stress, 
essentially an inverse viscosity
\begin{equation}
\dot{\varepsilon}_0 = \mu \left.{\partial \dot{\varepsilon}^{pl}_s \over \partial \sigma_s}
\right|_{\sigma_{flow}^+}
\label{gam0eq}
\end{equation}
In order to explore why such a change in $\dot\varepsilon_0$ might arise
we will now consider a somewhat simplified version of the theory
of viscoplasticity in amorphous solids developed in FL.

\subsection{Mesoscopic: viscoplasticity in amorphous solids}

In the model of viscoplasticity discussed in FL the plastic flow is
both rate and history dependent.  The history dependence of the model
enters through a set of state variables which describe the density of
``shear transformation zones''(STZ) of the type described in section
\ref{sec:molecobs}.  These STZ's are theorized to be essentially
two-state systems and are assumed to have a definite orientation.
That is to say that STZ's that are particularly susceptible to
deformation under one sense of shear may not be susceptible to
another, and when an STZ undergoes a transition it changes orientation
so as to be susceptible to an opposite applied shear stress.  For the
sake of simplification the STZ's are assumed to be either perfectly
aligned with the applied stress or anti-aligned with the applied
stress.  Furthermore, the first few non-linear terms which describe
the dynamics of these STZ's are conjectured.  These terms represent an
assumption that inelastic work done on the system may generate new
regions or eliminate existing regions.

In FL the rate of plastic strain is related to
the rate at which STZ's transform between their two states,
\begin{equation}
\label{epdot1frac}
\dot{\varepsilon}^{pl}_s= V_z\,\Delta\varepsilon\,\left[R_+\,n_
+-R_-\,n_-\right],
\end{equation}
where $V_z$ is the typical volume of a region, $\Delta\varepsilon$ is
the increment of local strain due to an individual transformation,
$n_\pm$ are the population densities of STZ's in each of the two
states, and $R_\pm$ are functions of the stress describing the rate at
which transitions occur between the two states.  The evolution equations of
$n_\pm$ are written in terms of a master equation\cite{Eyring36,Krausz75}
\begin{equation}
\label{ndot}
\dot n_{\pm}= R_{\mp}\,n_{\mp}-R_{\pm}\,n_{\pm} - 
C_1\,(\sigma_s\,\dot\varepsilon_s^{pl})\,n_{\pm} 
+C_2\,(\sigma_s\,\dot\varepsilon_s^{pl}).
\end{equation}
where $C_1$ and $C_2$ are constants associated with the non-linear
terms which determine the rate of STZ annihilation and creation.
The equations of motion have two steady states: a ``jammed'' or
``hardened'' state below the critical stress for plastic flow and a
flowing state above this stress.  For the flowing steady state
\begin{equation}
n_{\pm} = {C_2 \over C_1} \mp {1 \over V_z\,\Delta\varepsilon\,C_1\,\sigma_s}.
\label{yieldcrit}
\end{equation}

The specifics of the choice of the functions $R_\pm$ and their
dependence on the stress are important for determining the time
dependence of the plastic flow.  In FL the transition rates are
written as volume activated processes.  That is, the rates are written
in the form
\begin{equation}
\label{epdotfrac}
R_{\pm}=R_0 \,\exp\left[-{\Delta V^*(\pm\sigma_s)\over v_f} 
\right],
\end{equation}
where for the purpose of this analysis we will assume $R_0$ to be a
constant attempt frequency, $v_f$ is a free volume per particle, and
$\Delta V^*$ is a free volume needed to activate a transition.  The
volume needed to activate the transition $\Delta V^*$ is a function of
the applied shear stress which is chosen to have the simplest one
parameter functional form for which the volume is assured to be
non-negative.
\begin{equation}
\label{V0sigmafrac} \Delta V^*(\sigma_s)= V_0^*\,\exp(-\sigma_s/\bar\mu)
\end{equation} 
where $V_0^*$ is the free volume needed to activate a transition at
zero stress, and $\bar\mu$ is a modulus which characterizes the
sensitivity of the activation volume to the applied stress.  Note that
in general $V_0^* \gg v_f$ and these rates are negligible unless
$\sigma_s \approx +\bar\mu$.  Since we are considering the material
response around $\sigma_s = +\sigma_{flow}$, we are in a regime where
$R_+ \gg R_-$.

Taking this formulation of the transition rates into account, we can
consider the rate of deformation described by Eq. (\ref{epdot1frac}) in the
the steady-state flow regime of Eq. (\ref{yieldcrit}).
\begin{equation}
\dot{\varepsilon}^{pl}_s \;\approx\; V_z\,\Delta\varepsilon\,R_+\,n_
+ \;=\; {R_+ \over C_1}  [\sigma_{flow}^{-1} - \sigma_s^{-1}]
\label{floweq}
\end{equation}
where $\sigma_{flow} = (C_2\, V_z\,\Delta\varepsilon)^{-1}$.
So, we can evaluate $\dot{\varepsilon}_0$ in Eq. (\ref{gam0eq}) using
Eqs. (\ref{epdotfrac}-\ref{floweq}) to be
\begin{equation}
\dot{\varepsilon}_0 \;=\; {\mu \over C_1 \sigma_{flow}^2} R_0\,\exp\left[-{V_0^*\over v_f}
e^{-\sigma_{flow} / \bar\mu}\right]
\label{gam0eq2}
\end{equation}
This last equation provides a first clue as to which aspect of the
microscopic behavior is responsible for the observed change from
ductile to brittle failure.  First, note that we can reasonably
neglect the prefactors to the exponential since the effect of changes
in these terms will be substantially less dramatic.  Furthermore, the
ratio $V_0^*/v_f$, which was already noted to be a large number, is
expected to depend primarily on the relative sizes of the particles
which are the same in both systems.  Since the possibility of a
substantial change in $\sigma_{flow}$ has been ruled out, the only
remaining parameter in this expression is $\bar{\mu}$.  The double
exponential causes $\dot{\varepsilon}_0$ to be suppressed by a factor
of $\exp(-V_0^*/v_f)$ when $\bar\mu$ becomes large.  Moreover
Eq.~(\ref{gam0eq2}) is most sensitive to changes in $\bar\mu$ when
$\bar\mu \approx \sigma_{flow}$.  The investigations of analogous
amorphous systems via computer simulation in FL suggested that
$\bar\mu$ does indeed fall in this range.

The observation that $\dot\varepsilon_0$ is highly sensitive to
changes in $\bar\mu$ means that the sensitivity of the material flow
rate to a change in applied shear stress is highly dependent on the
sensitivity of the deformable regions (STZ's) in the solid.  Relating
this to Freund and Hutchinson's fracture model this implies that the
observed change from ductile to brittle failure seems to be due to a
corresponding change from ``floppier'' to ``stiffer'' weak regions in
the solid.  But while the viscoplasticity theory leads us to these
conclusions it does not elucidate how one might quantify these ideas
and relate them to the molecular potentials.

\subsection{Microscopic: simplified model of a two-state region}

\begin{figure}[tb]
\epsfxsize=5.0in
\centerline{\epsfbox{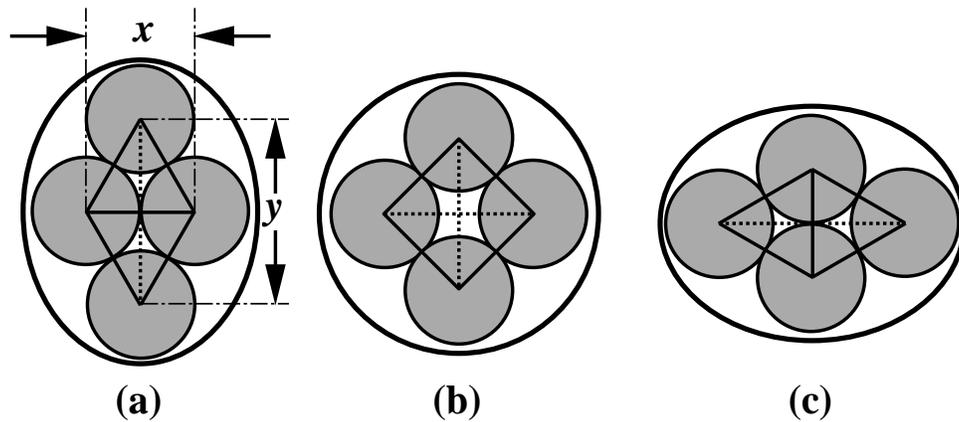}}
\caption{Diagram of four identical interacting particles making a
transition from one stable configuration to another.  The middle
configuration is the saddle-point configuration for this transition.
} \label{saddle}
\end{figure}

This idea of ``floppy'' or ``stiff'' STZ's can be made more meaningful
by considering a simplified model of molecular rearrangements.  The
model should be consistent with the observations of Argon, Spaepen and
co-workers, i.e. it should be capable of rearranging in a local
way,\cite{Spaepen77,Argon79a,Argon79b,Spaepen81,Argon83} and also with
the observations made by Srolovitz and co-workers, i.e. it should be a
``$\tau$-defect.''\cite{Srolovitz81,Srolovitz83} Figure~\ref{saddle}
shows the most stripped-down model of what these two-state systems
must look like on the molecular level, four molecules interacting via
a two-body interatomic potential.  Since this unit is embedded in the
solid it is constrained from undergoing translation or rotation.  For
particular choices of the interatomic potential this four-molecule
unit is inherently a two-state system.  That is to say that for a
Lennard-Jones or similar potential the energy is minimized by having
as many bonds near the equilibrium bond length as possible.  In this
system there are two degenerate ground states, illustrated in
Figs. \ref{saddle}(a) and \ref{saddle}(c), in which five of the six
bonds are of this length.

Because transitions between the two states of our four-body unit are
associated with the development of strain in the solid, the material
response must depend upon the rates at which these transitions occur.
At high temperatures the transition rates are dominated by rare
thermal events which occur only as $\exp(-\Delta U / kT)$, where
$\Delta U$ is the energy difference between the ground state and the
saddle point illustrated in Figure~\ref{saddle}(b).  This is exactly
the approach used to describe the time dependent strain in theories of
deformation kinetics such as those of Eyring, Spaepen and
Argon.\cite{Eyring36,Krausz75,Spaepen77,Argon83} Such a formalism is
not of use here.  In the present system $\Delta U$ is on order unity
while $kT$ is three orders of magnitude smaller.  Physically this
means transitions will be driven rather than thermally activated.
This is, of course, cause for alarm.  The statistical approach of the
theories of deformation kinetics in high temperature systems utilized
the statistical nature of the energetic fluctuations to discern a time
scale.  How can a statistical theory be developed when these fluctuations
are not relevant?  Instead, consider the solid to be composed of an
ensemble of these two-state systems, some small fraction of which are
close to a free volume induced transition.  By using this ensemble
picture it is possible to preserve the rare-event aspects of
transition state theory in order to extract a relevant time scale.

The concept of a free volume induced transition has been mentioned
here, but although this concept has been discussed in some detail in
FL, the molecular details of such a transition have not been
discussed.  In particular, it is necessary to be able to calculate
$\Delta V^*(\sigma_s)$, the shear stress dependent free volume needed
to activate the transition.  The following paragraphs will attempt to
answer the following question: For the model two-state system,
constrained by its surroundings to a certain area, what is the maximum
shear that it can support before being driven into a different state?
If this question can be answered, then, given some applied shear
stress, it will be possible to determine the free area (our
two-dimensional equivalent of free volume) at which a region will
become unstable.

With this picture in mind let us consider in some detail what is going
on physically.  We can parameterize the energy of the four-particle
system by only two parameters, its $x$ and $y$ dimensions, if we
constrain it from rotating, translating or deforming in an asymmetric
way.
\begin{equation}
{\cal U}(x,y) = U(x) + U(y) + 4 U({1 \over 2}\sqrt{x^2 + y^2})
\end{equation}
Here $U$ can be any two-particle potential, but we will concern
ourselves with $U_{LJ}$ and $U_{CLJ}$ described in section
\ref{sec:method}.  Furthermore, consider the case when the area
(two-dimensional volume) of the system remains constant by imposing the constraint $A=xy$.  We can define a local equivalent shear stress
\begin{equation}
\Sigma_s(x,y) = {1 \over 2}({1 \over y} {\partial {\cal U} \over \partial x} 
 - {1 \over x} {\partial {\cal U} \over \partial y}).
\label{taueq}
\end{equation}

At this point it is possible to understand why such two-state STZ's
would be visible as ``$\tau$-defects,'' the regions of anamolously high
shear stress described by Srolovitz, {\it et. al.}
\cite{Srolovitz81,Srolovitz83}  Consider the condition for the lowest energy
of the configuration,
\begin{equation}
{d{\cal U} \over ds} = {A \over \sqrt{x^2 + y^2}} 
({1 \over y} {\partial {\cal U} \over \partial x} 
 + {1 \over x} {\partial {\cal U} \over \partial y}) = 0,
\end{equation}
where the path of constraint is traversed by a unit speed curve
parameterized by $s$ such that $ds^2 = dx^2 + dy^2$.  We immediately
note that the condition for equilibrium is {\bf not} the same as the
condition for zero shear stress.  In general these two conditions are
not simultaneously satisfiable.  It is important to note
that an exception to this, i.e. a case in which the lowest energy
configuration has no shear stress, is the case where the molecules
interact only via nearest neighbor interactions.  This is particularly
interesting in light of the simulations since the limit where $\lambda
\gg 1$ in Eq. (\ref{CLJpot}) is the limit of solely nearest neighbor
interactions.  Therefore, we expect that the CLJ potential ($\lambda =
1.5$), which resulted in increased brittleness in the simulations, should
show lower levels of internal shear stresses and fewer
``$\tau$-defects'' than the LJ potential ($\lambda = 1$) which resulted in
increased ductility.  Thus, the microscopic model
strongly suggests that the range of the intermolecular potential is
crucial in determining whether these STZ's are visible as
``$\tau$-defects.''.

\begin{figure}[tb]
\epsfxsize=5.0in
\centerline{\epsfbox{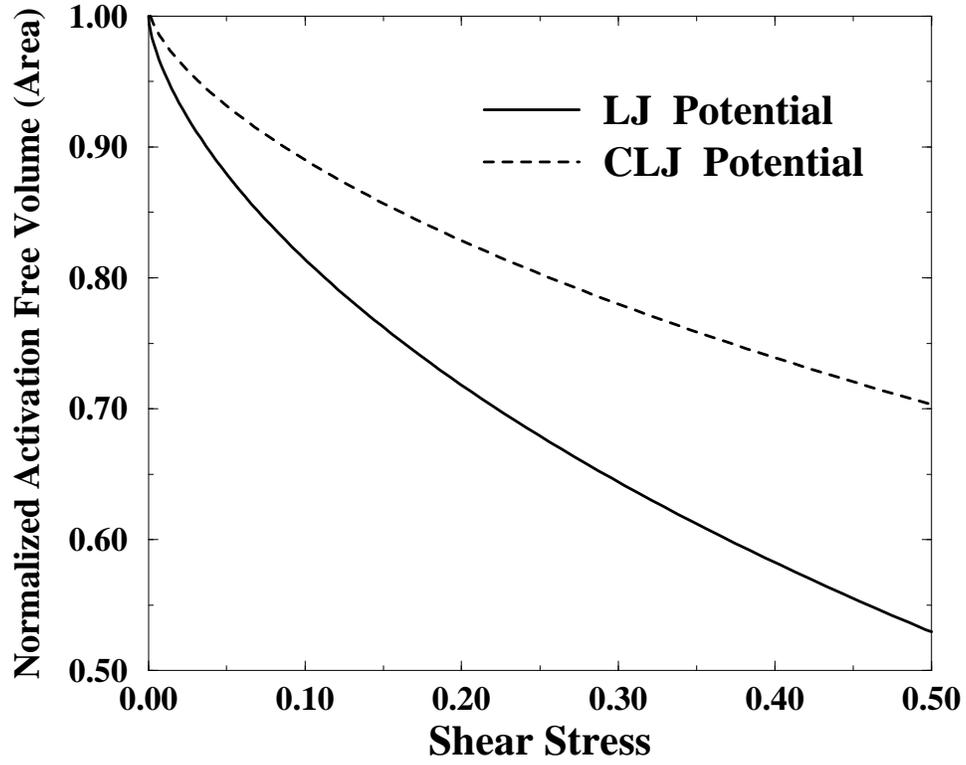}}
\caption{The activation free volume (area) at shear stress $\Sigma_s$ 
divided by the activation free volume (area) at zero applied shear
stress, $A_f(\Sigma_s)/A_f(0)$.  The activation free volume (area)
corresponds to the excess volume at which the four particle system
illustrated in Figure~\ref{saddle} becomes unstable to the applied
shear.  Stresses are given in units of $e/d_0^2$.} \label{instability}
\end{figure}

We return now to the question of when the two-state STZ will become
unstable to an externally applied shear stress.  The condition for
instability can be written
\begin{equation}
{d\Sigma_s \over ds} = {1 \over \sqrt{x^2+y^2}}(x {\partial \Sigma_s \over \partial x}- y {\partial \Sigma_s \over \partial y})\;=\;0
\end{equation}
where we have again traversed the path of constraint by a unit speed
curve.  We can now define the equivalent of a free volume in our
system, $A_f = A - A_0$.  Here $A_0$ is the equilibrium area of the
four-body system at zero applied shear stress $\approx
\sqrt{3}\:(d_0^2)$.  Figure~\ref{instability} shows the value of
$A_f(\Sigma_s)/ A_f(0)$ at which instability sets in for values of the
shear stress $\Sigma_s$.  In order to illustrate the suppression of
$\Delta V^*$, and by analogy $A_f$, as the stress is raised to the
flow stress, Figure~\ref{instability} spans a range of shear
stresses up to our observed flow stresses in the fracture samples.
This graph looks suggestively similar to the form guessed in
Eq.(\ref{V0sigmafrac}).  In actual fact, however, in the vicinity of $\Sigma_s
= 0$, $A_f$ is a power law and not an exponential decay.  This suppression of the activation volume can be related to values for
$\bar\mu$ in FL.  The activation
area at $\Sigma_s=0.4$ is $58\%$ of its value at zero in the LJ case
and $74\%$ in the CLJ case; this corresponds to
$\bar\mu_{LJ}=0.73$ and $\bar\mu_{CLJ}=1.3$.  Thus, longer range
intermolecular potentials correspond to a solid with ``floppier''
two-state regions.  This result is significant since longer range
potentials also implied larger local stresses, and the existence of
``$\tau$-defects.''  As expected from the previous analysis, the
toy model with the CLJ potential has a higher value of $\bar\mu$ and,
therefore, corresponds to a solid with ``stiffer'' two-state regions.
This is in keeping with our expectations since ``stiffer'' two-state
regions should also correspond to a lower value of $\dot\varepsilon_0$
and, therefore, by Freund and Hutchinson's model to a lower minimum
energy release rate for brittle fracture from Eq. (\ref{Geq}).  The
CLJ solid is observed to undergo brittle fracture.

The analysis presented here is clearly only a first step toward a
rigorous first-principles theory of brittle-ductile transitions in
noncrystalline solids.  Future investigations will hopefully allow
more explicit connections to be made between the molecular level
structures which are quantifiable via diagnostics such as $D^2_{min}$
and the observed fracture behavior.  Progress will require further
developments in our understanding of the molecular physics of
deformation in non-crystalline solids.

\section*{Acknowledgments}

I would like to acknowledge J.S. Langer for his guidance and
encouragement, Alexander Lobkovsky for sharing his ideas and results
regarding decohesion models, and A.S. Argon, B.L. Holian, M. Marder and
R.L.B. Selinger for helpful discussions.  This work was supported by
the DOE Computational Sciences Graduate Fellowship Program and DOE
Grant No. DE-FG03-84ER45108.  The work was also supported in part by
National Science Foundation grant CDA96-01954, Silicon Graphics Inc.,
and the Cornell Theory Center.

\bibliography{biblio/MDcracks}

\end{document}